\documentclass[prb,twocolumn,amsfonts,amssymb,amsmath,floatfix,tightenlines,superscriptaddress,nofootinbib,prbbib-new,showkeys]{revtex4-2}

\usepackage{bm}

\def\pair{\textrm{pair}}
\def\coh{\textrm{coh}}

\newcommand{\LSCO}{La$_{2-x}$Sr$_x$CuO$_4$}

\newcommand{\YBCO}{YBa$_2$Cu$_3$O$_{6+\delta}$}

\newcommand{\HgBCO}{HgBa$_{2}$CuO$_{4+\delta}$}
\newcommand{\BISCO}{Bi$_2$Sr$_{2}$CaCu$_2$O$_{8+\delta}$}

\newcommand{\vect}[1] {\mathbf{#1}}

\newcommand{\uk}{u_{\mathbf{k}}}
\newcommand{\vk}{v_{\mathbf{k}}}
\newcommand{\xicoh}{\xi_0^{\mathrm{coh}}}
\newcommand{\mupair}{\mu_{\mathrm{pair}}}

\newcommand{\kF}{k_{\mathrm{F}}}
\newcommand{\EF}{E_{\mathrm{F}}}
\newcommand{\Tc}{T_{\mathrm{c}}}
\newcommand{\Hc}{H_{\mathrm{c2}}}
\newcommand{\kB}{k_{\mathrm{B}}}
\newcommand{\vF}{v_{\mathrm{F}}}
\newcommand{\vecq}{\mathbf{q}}
\newcommand{\veck}{\mathbf{k}}

\usepackage{amsfonts,amssymb}
\usepackage[subnum]{cases}
\usepackage{mathrsfs}
\usepackage[tone]{tipa}
\usepackage{amsmath}
\usepackage{graphicx}
\usepackage{footmisc}
\usepackage[colorlinks=true,citecolor=blue,linkcolor=blue,urlcolor=blue]{hyperref}
\usepackage{bm}          
\usepackage{natbib}
\usepackage{soul} 
\usepackage{color}
\usepackage{longtable}
\usepackage{ textcomp }
\usepackage{verbatim}
\usepackage[english]{babel}
\usepackage{times}

\begin{document}

\title{Test for BCS-BEC Crossover in the Cuprate Superconductors}

\author{Qijin Chen}
\email[Emails: ]{qjc@ustc.edu.cn; levin@jfi.uchicago.edu}
\affiliation{Hefei National Research Center for Physical Sciences at the Microscale and School of Physical Sciences, University of Science and Technology of China,  Hefei, Anhui 230026, China}
\affiliation{Shanghai Research Center for Quantum Science and CAS Center for Excellence in Quantum Information and Quantum Physics, University of  Science and Technology of China, Shanghai 201315, China}
\affiliation{Hefei National Laboratory, University of  Science and Technology of China, Hefei 230088, China}
\author{Zhiqiang Wang}
\affiliation{Department of Physics and James Franck Institute, University of Chicago, Chicago, Illinois 60637, USA}
\author{Rufus Boyack}
\affiliation{Department of Physics and Astronomy, Dartmouth College, Hanover, New Hampshire 03755, USA}
\author{K. Levin}
\email[Emails: ]{qjc@ustc.edu.cn; levin@jfi.uchicago.edu}
\affiliation{Department of Physics and James Franck Institute, University of Chicago, Chicago, Illinois 60637, USA}

\date{\today}

\begin{abstract}
In this paper we
address the question of whether high-temperature superconductors have anything
in common with BCS-BEC crossover theory.
Towards this goal, we present a proposal and related predictions which provide a concrete test for the applicability of this theoretical framework. 
These predictions
characterize the behavior of the Ginzburg-Landau coherence length,
$\xicoh$, near the transition temperature $\Tc$, and
across the entire superconducting $\Tc$ dome in the phase diagram.
That we are lacking a systematic characterization of $\xicoh$ in the
entire class of cuprate superconductors is perhaps surprising, as it
is one of the most fundamental properties of any superconductor.  This
paper is written to motivate further experiments and, thus, address
this shortcoming.  Here we show how measurements of
$\xicoh$ contain direct indications for whether or not the
cuprates are associated with BCS-BEC crossover and, if so, where
within the crossover spectrum a particular superconductor lies.
\end{abstract}


\maketitle

\section{Introduction}

The subject of high-temperature superconductivity in the cuprates is
by now a mature field with a diverse array of candidate theories.
This applies as well to theories of the mysterious pseudogap which has
captured the attention of the community.  It is notable that there is
still no consensus about the nature of the machinery and the mechanism
behind this phenomenon. This should be clear from the large
number of review articles~\cite{Lee2006,Chen2005,Keimer2015,Fradkin2015}, which
represent a range of different perspectives and viewpoints.  It can be
plausibly argued that, as the field is so mature, what is most
needed now is for candidate cuprate theories to formulate testable,
preferably falsifiable predictions.

This is the goal of the present paper for one particular theory,
called `BCS-BEC crossover theory'.  Here we address the question of
whether high-temperature superconductors have anything in common with
BCS-BEC crossover theory. We understand this to be a highly
controversial issue, but note that this subject has received recent
attention in the literature~\cite{Harrison2022,Sous2023}.  For this
purpose, we focus on the Ginzburg-Landau (GL) coherence length,
$\xicoh$, appropriate to near or slightly above $\Tc$. In
this paper we provide direct predictions for its behavior 
as a function of hole doping.  
We argue that these predictions can be used to directly assess the
appropriateness of BCS-BEC crossover theory for the cuprate family,
when $\xicoh$ is measured systematically across the $\Tc$ dome in the phase
diagram and for the different cuprate families.

This crossover theory has the advantage over many other cuprate theory
candidates of potentially being applicable to a wide collection of
strongly correlated superconductors.  This broad range of
applicability is exploited in the present work.  Over the years we
have acquired a knowledge base which has shown how to connect
different types of experimental findings with `crossover physics'.
BCS-BEC crossover candidate materials include iron-based
superconductors~\cite{Kasahara2016,Kasahara2014,Okazaki2014,Mizukami2021,Hanaguri2019,Shibauchi2020,Kang2020,Faeth2021},
organic superconductors~\cite{Mckenzie1997,Imajo2021,Matsumura2022,Oike2017,Suzuki2022},
magic-angle twisted bilayer (MATBG)~\cite{Cao2018,Oh2021} and
trilayer graphene (MATTG)~\cite{Park2021,Kim2022}, gate-controlled
two-dimensional devices~\cite{Nakagawa2021, Saito2016, Nakagawa2018},
interfacial superconductivity~\cite{Richter2013,Bozovic2020,Cheng2015}, and
magneto excitonic condensates in graphene
heterostructures~\cite{Liu2022}.

The theory of BCS-BEC crossover~\cite{Leggett1980,Eagles1969,Nozieres1985,Chen2005,Giorgini2008,Randeria2014,Chen2022} belongs to a class of preformed-pair theories
associated with relatively strong `pairing glue'.
As a result, fermion pairs form at a higher temperature before they Bose condense at the superfluid transition temperature $\Tc$, as found in the BEC phase of a Bose superfluid.
Importantly, there is a continuous evolution between the two endpoints of a crossover theory: the conventional, weak-pairing BCS limit and the strong-pairing BEC limit.
We emphasize that this theory pertains only to the machinery of superconductivity.
It calls for a revision of the more familiar BCS approach, while still contemplating
a charge $2e$ pairing-based scheme. It does not address the
specific microscopic pairing mechanism.

Whether crossover theory is applicable to the cuprates or not is the
question we wish to help address in the course of future experiments.
That the transition temperature is high and there are indications that
the GL coherence length (in comparison to traditional
superconductors) is small are argued~\cite{Leggett2006} to be
suggestive of strong pairing `glue'. But it is clearly of interest to
find more definitive and quantitative evidence for or against this scenario.

Here and in the vast literature on ultracold atomic Fermi
gases~\cite{Chen2005,Giorgini2008,Randeria2014}, a
superconductor/superfluid in the `crossover' regime is conventionally
viewed as belonging somewhere intermediate between BCS and
BEC.  There should be little doubt that the cuprates are not in the
BEC limit.  In this regime, all signs of fermionic physics have
disappeared, which is clearly not the case for the cuprate
superconductors; this point has been made recently by Sous \textit{et. al.}~\cite{Sous2023} in their analysis of the behavior of the fermionic
chemical potential.  The more relevant issue is whether the
high-temperature superconductors can be described as belonging to an
intermediate regime, somewhere between BCS and BEC and, if so, where
in this spectrum  a given cuprate might lie.

Indeed, even when a superconductor is on the fermionic side of the
crossover, it can behave in a rather anomalous fashion both above and
below $\Tc$. We list here three necessary conditions for this
crossover scenario to apply.  (i) It is associated with the presence
of a fermionic excitation gap or `pseudogap' which has a temperature
onset, $T^*$, substantially above $T_{\mathrm{c}}$ (say, $T^*/\Tc \gtrsim
1.2$). (ii) It is also associated with sizable ratios of the
zero-temperature gap to Fermi energy, $\Delta_0/E_{\text{F}}$, (say,
$\Delta_0/E_{\mathrm{F}} \gtrsim 0.1$) and finally (iii) it has an anomalously
small GL coherence length (say, $k_{\mathrm{F}} \xi_0^{\coh} \lesssim 30$, where
$k_{\mathrm{F}}$ represents the ideal-gas measure of the carrier density).

In this context, it should be noted that large $T^*/\Tc$ is a
necessary but not sufficient criterion for a pairing pseudogap, as
there are alternative reasons why this ratio might be
large~\cite{Chand2012}.  On the other hand, a moderately large
$\Delta_0/\EF$ may be more indicative of BCS-BEC crossover, but this
ratio can be rather complicated to assess. This is because $\EF$ is
usually hard to quantify in a typical superconductor, as it is related
to complex band structures. And for the cuprates one would presumably
have to quantify this ratio over the entire $\Tc$ dome as a function
of hole doping.

This leaves the GL coherence length as arguably the most useful
parameter for characterizing BCS-BEC crossover.  This length scale, which
essentially reflects normal-state pairing correlations, should not be
confused with other length scales such as the London penetration
depth, which characterizes the superconducting components of the system. 
To be specific, the zero-temperature London penetration depth is related 
to the density-to-mass ratio of the constituent fermions, which is
independent of pairing correlations (here, for simplicity, we consider a 3D superconductor in free space).  
We also emphasize here that in the crossover regime, this
coherence length deviates from its BCS-limit expression and that it is
similarly distinct from a measure of the size of the Cooper pairs.

\begin{figure*}[htp]
\centering
\includegraphics[width=6.1in]{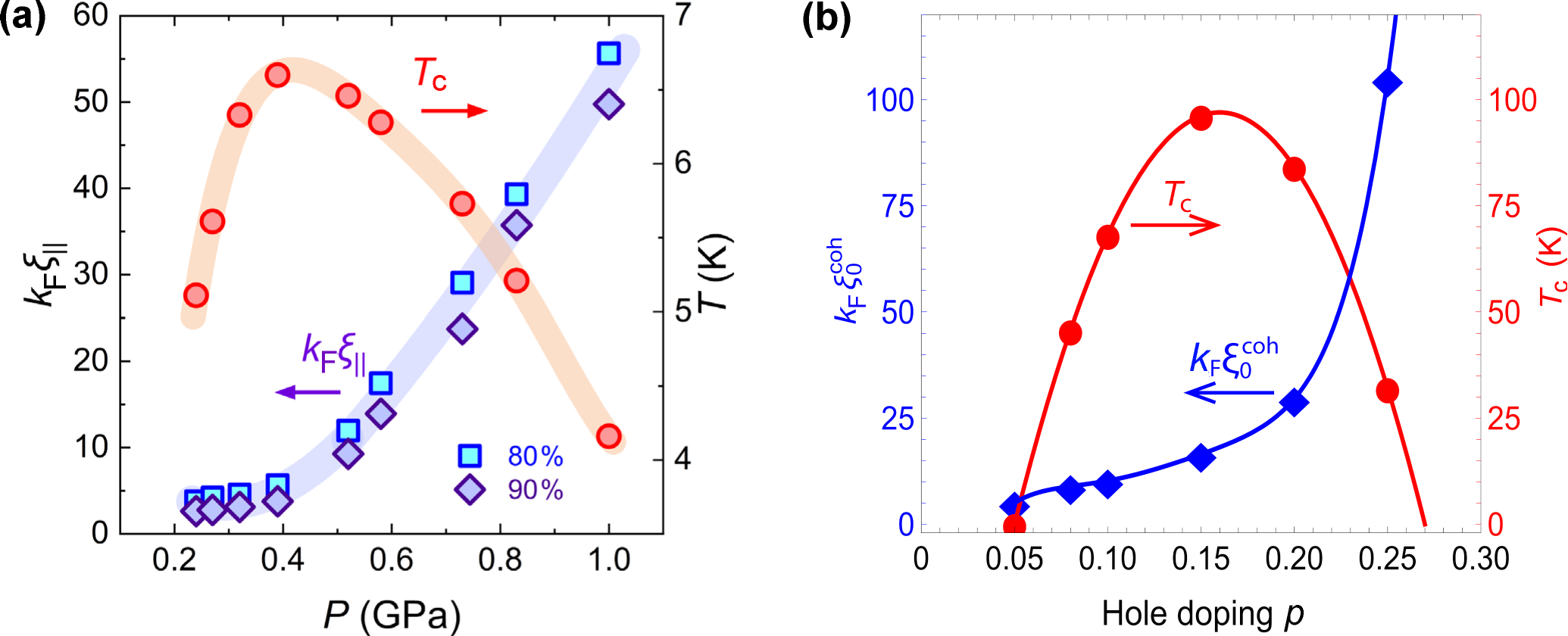}
\caption{\textbf{Comparison between experiment and theory for the in-plane coherence length and the $\Tc$ dome.} (a) Pressure dependence of the measured
in-plane coherence length $\kF \xicoh$ near $\Tc$, and superconducting
transition temperatures in $\kappa$-(BEDT-TTF)$_4$Hg$_{2.89}$Br$_8$, taken from Suzuki \textit{et. al.}~\cite{Suzuki2022}.
Here $\kF$ is determined from the carrier density measured by the Hall coefficient.
The $\Tc$ dome with overlain coherence length provides a rather ideal prototype for BCS-BEC crossover physics.
(b) Calculated in-plane Ginzburg-Landau coherence length, based on fits to the cuprate
phase diagram in Fig.~\ref{fig:ExptPhaseDiag}. This coherence length should be associated with measurements at very low magnetic fields and near $T \approx \Tc$. The red circles indicate the selected hole concentrations on the
$\Tc \sim p$ dome where both $T^*$ and $\Tc$ were simultaneously fitted to yield
the computed coherence lengths (blue diamonds).}
\label{fig:1}
\end{figure*}

\section{Results}
\subsection{Coherence length in BCS-BEC crossover}
A recent paper~\cite{Suzuki2022} on a candidate organic
superconductor has provided a template of the coherence length for us
to use here in presenting predictions for the cuprates.  This is shown
in Fig.~\ref{fig:1}(a) where the dimensionless coherence length
$\kF \xi_0^{\coh}$ is plotted across the entire $\Tc$ dome.
Here the nominal Fermi momentum $\kF$ simply reflects the carrier density. 
For this particular organic superconductor, pressure is used as a
tuning parameter to effect the crossover between the weak-coupling and
strong-pairing limit.

A central result of the present paper is establishing the counterpart 
behavior of Fig.~\ref{fig:1}(a) for the cuprates, particularly for the entire range of hole doping over the $\Tc$ dome. 
This is shown in
Fig.~\ref{fig:1}(b).
Indeed, the GL coherence length has become a preferred quantity to measure for many of the newer BCS-BEC candidate systems~\cite{Park2021,Nakagawa2021}. 

The coherence length that we are interested in here
can be obtained in several different ways. In principle, it
enters into the slope of the upper critical magnetic
field, $\Hc$, very near $\Tc$: 
\begin{equation} 
\frac{d\Hc}{dT} \bigg\vert_{T=\Tc}= -\frac{\Phi_0}{2 \pi \Tc (\xi_0^{\text{coh}})^2}
\quad\textrm{with}\quad
\Phi_0 =\frac{hc}{|2e|}.
\nonumber
\end{equation}
This is based on using the temperature-dependent coherence length $\xicoh(T)$, which
is defined in terms of the quantity of interest, $\xicoh$,
as $\xicoh(T)=\xicoh /\sqrt{(\Tc -T)/\Tc}$ as in conventional superconducting fluctuation theories. 
As discussed in the context of MATTG~\cite{Park2021}, extracting $\xicoh$ experimentally from $d\Hc/dT$ is not entirely straightforward as it involves determining $\Tc(H)$ in the presence of a substantial field-induced broadening of the transition.

Alternatively, in line with the philosophy in this paper,
one can avoid some of these complications by
determining the GL coherence length 
through studies of the fluctuation
magnetotransport~\cite{Suzuki1991} in the normal state above $\Tc$. Such experiments are generally performed in combination
with theoretical analyses based on the Aslamazov-Larkin (AL) pairing fluctuations~\cite{VarlamovBook,Varlamov2018}.

\subsection{Theory overview}
We next evaluate the GL coherence length within 
BCS-BEC crossover theory.
In the presence of a vector potential, the effects from non-condensed pairs (and the associated pseudogap)
are inhomogeneous, and thus directly evaluating $\Hc(T)$ to extract
$\xi_0^{\coh}$ poses a great challenge to theory. 
By contrast, here we deduce the coherence length alternatively based on (normal-state) fluctuation theory~\cite{Patton1971,Ullah1991}.
Superconducting fluctuations are generally associated with
AL contributions~\cite{VarlamovBook} which reflect bosonic or pairing degrees of freedom.
Their contributions~\cite{VarlamovBook} to transport and thermodynamics generally scale as powers of 
$\epsilon \equiv (T- \Tc)/\Tc $ or the effective chemical potential of the pairs.

There is a rather direct association between the AL treatment of conventional weak-pairing
fluctuations and that deriving from the
strong-pairing regime.  The conventional fluctuation propagator~\cite{VarlamovBook}
 depends on two parameters, $\epsilon$ and $\xi_0^{\coh}$. Similarly, for strong pairing
the pair propagator, called the $t$-matrix, depends on an analogous pair of parameters, 
the pair chemical potential $\mu_\pair$ and the inverse pair mass $M_\pair^{-1}$.
While conventional fluctuation transport calculations are complex~\cite{VarlamovBook},
the central parameters $\epsilon$ and $\xi_0^{\coh}$ 
are essentially all that is needed to arrive at the entire collection 
of transport coefficients. Importantly, those calculations serve as a template for doing
transport in the strong-pairing regime~\cite{Boyack2018,Boyack2019}, 
provided one makes the association $ \epsilon \rightarrow |\mu_\pair|/\Tc$
and similarly relates the pair mass $M_\pair$ to the coherence length $\xicoh$ within the strong-pairing theory via
\begin{equation}
\hbar^2/[2M_\pair(T_\text{c})(\xi_0^{\coh} )^2]= \kB \Tc.
\label{eq:4}
\end{equation}

It should not be surprising then that (because BCS theory and its BCS-BEC crossover
extension treat the Cooper
pair degrees of freedom as quasi-ideal bosons interacting
indirectly only via the constituent fermions),
the expression for the transition temperature $\Tc$ essentially follows that of an ideal
Bose gas (see Methods). For three dimensions (3D), this is given by
\begin{equation}
T_\text{c} = \left(\frac{2\pi}{\mathcal{C}}\right)
\left[ \frac{\hbar^2}{\kB}  \frac{n_\pair^{2/3}(T_\text{c})} { M_\pair(T_\text{c})}\right],
\label{eq:2}
\end{equation}
where $\mathcal{C} =\left[\zeta(3/2)\right]^{2/3}$ with $\zeta(s)$ the Riemann zeta function.
In this equation, $n_\pair$ and $M_\pair$ represent the respective number density and mass of the preformed Cooper pairs,
which will condense at the transition. These parameters must be determined self consistently (see Methods).

It then follows from
Eqs.~\eqref{eq:4} 
and \eqref{eq:2}
that $\xi_0^{\coh}$ assumes a very simple form;
it depends only on the non-condensed or normal-state
pair density $n_\pair$ presumed at the onset of the
transition: 
\begin{equation}
\kF \xi_0^{\text{coh}}= 1.2 (n/n_\pair)^{1/3},
\label{eq:5}
\end{equation}
where $\kF^3$ reflects the total particle density $n$.

It should be noted that the above discussion can be extended to 2D as well, leading to a similar
conclusion for the GL coherence length~\cite{Chen2022}:
\begin{equation}
\kF \xi_0^{\text{coh}} = 1.6 (n/n_\pair)^{1/2}.
\label{eq:6}
\end{equation}
For the quasi-2D cuprates, both $M_{\pair}$ and $\xi_0^{\coh}$ in Eq.~\eqref{eq:4} are naturally anisotropic, but here we
are interested in the in-plane coherence length so that, as in experiment, only the in-plane parameters
will be used throughout.

We note that the above equations are easy to understand physically. The GL coherence length is a length
representing the effective separation between preformed pairs. It relates to the density of pairs 
at $\Tc$ as distinct from the
pair size.
In BCS theory there are almost no pairs present at $\Tc$ and the length which represents their
average separation is necessarily very long. As pairing becomes stronger more pairs form and their
separation becomes shorter. On a lattice, in the BEC regime their separation is bounded from below by the
characteristic lattice spacing and
$\xi_0^{\coh}$ approaches an asymptote set by the inter-particle distance as the system varies from BCS to BEC.

More importantly, the rather natural expressions for 
$\kF \xi_0^{\text{coh}}$ in Eqs.~\eqref{eq:5} and \eqref{eq:6}
also reveal the location of a given system within the BCS-BEC crossover. 
Since the number of pairs at $\Tc$ varies from approximately 0 in the BCS limit to $n/2$ in the BEC case, the GL
coherence length provides a quantitative measure of where a given superconductor is within the BCS-BEC
spectrum.

\subsection{Application to the cuprates}
\label{sec:application}

\begin{figure}[h]
\centering
\includegraphics[width=3.0in]{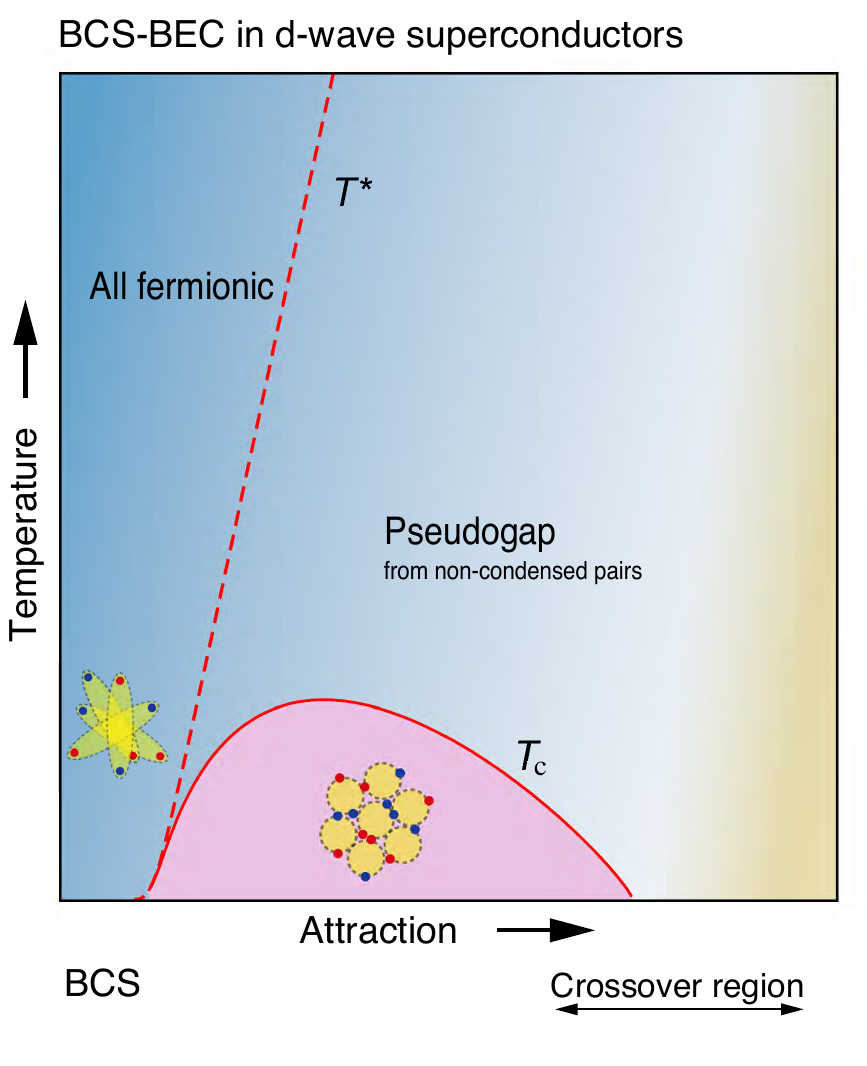}
\caption{\textbf{BCS-BEC crossover phase diagram for a $d$-wave superconductor with constant carrier density.} This diagram~\cite{Chen1999} shows that the system (with a single electronic conduction band nearly half-filled) has vanishing $\Tc$ before the onset of the BEC regime, where the zero-temperature fermionic chemical potential drops below the band bottom. This can be compared with the low-density $s$-wave case in which the BEC regime is in principle accessible.}
\label{fig:PhaseDiagram}
\end{figure}

In application to the cuprates
it is useful first to present a $\Tc$ versus attraction strength $|U|$ phase diagram for the case of $d$-wave pairing symmetry.
This is deduced~\cite{Kadanoff1961,Chen2005,Chen2022} based on Eq.~\eqref{eq:2} 
for $\Tc$ [see Methods]. 
Here, for the pseudogap onset temperature $T^*$ we use a straightforward mean-field theory.
The results are shown in Fig.~\ref{fig:PhaseDiagram}.
What is notable here is the fact that, in contrast to the
$s$-wave pairing in BCS-BEC crossover~\cite{Chen2022}, for the case of a $d$-wave superconducting order parameter,
the BEC regime is not generally accessible except when the underlying conduction band 
has an extremely low filling. Although not relevant to the cuprates which are near half-filling, there may be
other BCS-BEC crossover candidate systems which exhibit a
$d$-wave BEC phase. 

Heuristically, we understand the above contrast between $s-$ and $d-$wave pairing as a consequence of the fact that $d$-wave pairs are more extended in size, so that multiple lattice sites are involved in the pairing. Consequently, repulsion between pairs is enhanced due to a stronger Pauli exclusion effect experienced by these extended
pairs, and as a result their hopping is greatly impeded. Adding to this is the well known~\cite{Nozieres1985} 
observation that hopping of pairs on a lattice becomes more problematic in the strong-attraction regime,
since the paired fermions have to unbind in the process. While in a low carrier density, $s$-wave pairing superconductor,
$\Tc$ consequently approaches zero asymptotically in the BEC regime, generally
for $d$-wave superconductors, $\Tc$ will vanish before the BEC limit is reached.

The above discussion brings us to the central topic of this paper:
how one should determine whether the cuprates are associated with a BCS-BEC
scenario and, if so, where a given cuprate precisely lies in the spectrum of
BCS to BEC. Our proposal to quantitatively address this question is to
focus on the calculated GL coherence length, with the goal of providing a counterpart plot like that in Fig.~\ref{fig:1}(a), but now for the cuprates.

\begin{figure*}[htp]
\centering
\includegraphics[width=4.8in]{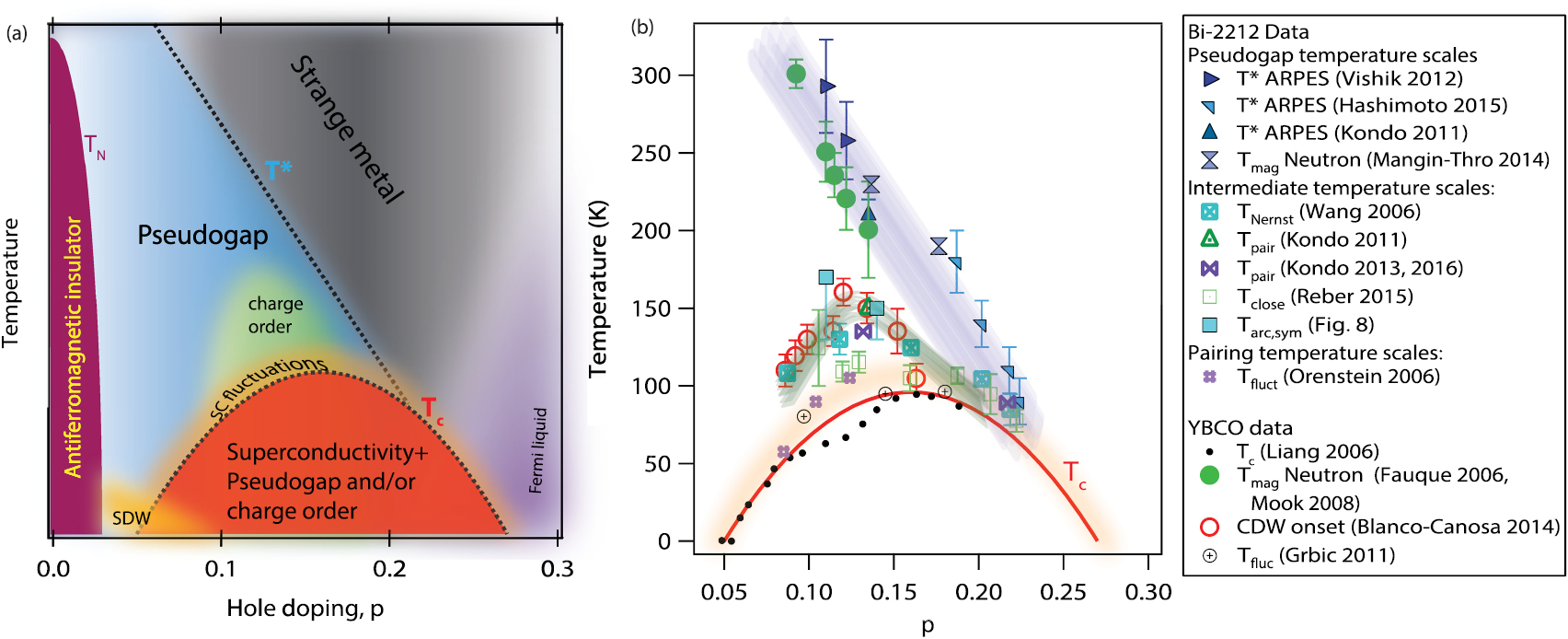}
\caption{\textbf{Experimental phase diagram for hole-doped cuprates, taken from Ref.~\onlinecite{Vishik2018}.} $T^*$ and $\Tc$ shown in (a)
are quantitatively plotted in (b). The error bars in (b) represent the standard deviation.}
\label{fig:ExptPhaseDiag}
\end{figure*}

To that end, the first immediate task is to connect the $d$-wave crossover phase diagram in
Fig.~\ref{fig:PhaseDiagram} with the experimental cuprate phase diagram in Fig.~\ref{fig:ExptPhaseDiag},
where the horizontal axis is hole doping $p$, instead of $|U|$. 
To establish the connection, we fit the calculated $T^*$ and $\Tc$ at a number of hole concentrations in the theory phase diagram
to their corresponding experimental values, and deduce the associated properties of the GL coherence length.
What is subtle but important here is that the phase diagram of Fig.~\ref{fig:PhaseDiagram} was
obtained for a fixed carrier density. For application to the cuprates we need to readjust the density at each
point in the $\Tc \sim p$ dome.

To be specific, by taking the experimental $T^*$, $\Tc$, and the corresponding density as input
fitting parameters from Fig.~\ref{fig:ExptPhaseDiag}, we can establish from $T^*/\Tc$ the magnitude of the attractive interaction ratio $|U|/t$, using the theoretical phase diagram in Fig.~\ref{fig:PhaseDiagram}.
Here $t$ is the effective hopping parameter.
Then fitting the numerical value of $T^*$ yields the value of $t$,
which determines the bandwidth and Fermi energy for each cuprate with a
different hole concentration. From the fitted parameters $\{ U, t\}$ and the hole concentration, we can
compute (see Methods) $n_{\pair}$ and $M_{\pair}$, using our $t$-matrix theory~\cite{Kadanoff1961,Chen1999}, and then extract the coherence length.

For definiteness, we adopt a quasi-2D band structure considered to be appropriate for the cuprates: 
$\epsilon_{\vect k} = (4t+4 t^\prime +2 t_z) - 2t (\cos k_x +\cos k_y) - 4 t^\prime \cos k_x \cos k_y - 2 t_z \cos k_z$ with $t^\prime/t = -0.3$. We presume a very small $t_z/t = 0.01$ is also present, but it should be stressed that
$\Tc$ has only a very weak logarithmic dependence on $t_z$~\cite{Chen1999}.
This band structure has a van Hove singularity which is prominent for the band fillings we address.

The predicted results for the GL coherence length based on our fitting procedure and BCS-BEC
crossover theory are presented in Fig.~\ref{fig:1}(b). These results show that, not unexpectedly, the coherence length
is predicted to decrease monotonically with increased underdoping of holes, reflecting that
the pairing strength is strongest in the most underdoped systems. 
Note that for the cuprates, the predicted minimum value of the coherence length is
not particularly short. This more moderate value for $\xi_0^{\coh}$
in the underdoped regime is associated with the $d$-wave symmetry of the cuprates. In this doping regime, $n_\pair$, the number of
pairs at the transition temperature $\Tc$, remains far below its maximum possible value of $n/2$;
stated alternatively,
the corresponding $|U|/t$ at these hole concentrations is smaller than the value of $|U|/t$ where $\Tc$ vanishes (see Fig.~\ref{fig:PhaseDiagram}).
This implies that the underdoped cuprates are still well within the fermionic side of the crossover `transition',
which is defined as where $\mu=0$ at $\Tc$. 

On the experimental side, in the earlier literature there is a
prototypical set of experiments~\cite{Suzuki1991} which address
$\xi_0^{\coh}$ in the immediate vicinity of the
transition. Importantly, this analysis is based on a
normal-state fluctuation analysis; as in a similar spirit to
the theoretical calculation of $\xi_0^{\coh}$, this avoids difficulties
associated with evaluating $d\Hc /dT \big\vert_{T=\Tc}$ more
directly.  As seen in Fig.~14(a) of Ref.~\cite{Suzuki1991}, this
analysis finds that in \LSCO\ single-crystal films, there is a rather
weak decrease of $\xicoh$ observed with increased
underdoping. However, in the overdoped regime the measured coherence
length is not as large as suggested in our Fig.~\ref{fig:1}(b).

This and related research have emphasized that experiments based on
standard fluctuation analyses below $\Tc$ are more
problematic than above $\Tc$. It is the shortness of the coherence length
itself which is causing the difficulty. More specifically, the short
coherence length results in a small characteristic energy associated
with vortex pinning centers.  This allows their motion to be more
readily thermally activated.  As a result, this enhanced vortex
depinning significantly increases the width of the resistivity
transition, making it difficult to determine the precise value of
$\Tc(H)$ and, similarly, $\xi_0^{\coh}$.

These $ T \approx \Tc$ studies which we focus on here should be
contrasted with coherence-length measurements at low temperatures
where use is made of the vortex core
size~\cite{Wen2003,Sonier2007}. Interestingly, here and in related
transport experiments~\cite{Ando2002} there are similar challenges in
measuring the coherence length which were attributed to the presence
of a vortex liquid rather than vortex solid phase.

There are also other potential complications stemming from
Fermi-surface reconstruction~\cite{Chan2020}, which can be viewed as
deriving from ordering in the particle-hole channel, seen at high
magnetic fields $H$.  If this reconstruction persists in the very low
$H$ limit, those regions of the $T$-$p$ phase diagram where
reconstruction appears will complicate the interpretation of $\Tc(H)$
and, in turn, affect the inferred $\xi_0^{\coh}$.  Indeed, it is now
understood that three cuprate families (\YBCO, \LSCO, and \HgBCO) each
show significant Fermi-surface reconstruction in magnetic
fields. These lead to non-monotonicity in the
inferred~\cite{Chan2020,Badoux2016,Wang2006} $\Hc(T=0)$ and related
$T=0$ coherence length~\cite{Sonier2007}, as a function of hole
doping.  We note that for the \BISCO~family, by contrast, it appears from 
Nernst measurements that $\Hc(T)$ may not have these dramatic
non-monotonicities in hole doping~\cite{Wang2003,Chang2012}, from
which one might presume that this family is not subject to Fermi
surface reconstruction.  Thus, these cuprates might be more suitable candidates for future experiments. 

\subsection{Fluctuation Temperature Scale in Cuprates}

There is another temperature scale besides $T^*$ and $\Tc$
apparent in the phase diagram of Fig.~\ref{fig:ExptPhaseDiag} which, 
for completeness, needs to be addressed within the crossover scenario.
We interpret this additional temperature scale in 
Fig.~\ref{fig:ExptPhaseDiag}
as~\cite{Chen2022} the onset temperature for superconducting fluctuations which, as a function of hole
doping in the cuprates, is observed to follow $\Tc$, although remaining well separated.

It should be clear that within the crossover scheme
the cuprates cannot be described by conventional fluctuation theory
owing to the existence of a pairing gap onset temperature, $T^*$, significantly higher than $\Tc$.
That is, in the presence of a pseudogap associated with preformed pairs, the
pairs are present over a much wider temperature range than in conventional fluctuation theory.

More specifically, as in fluctuation theories~\citep{VarlamovBook,Varlamov2018}, fluctuation contributions in the crossover scenario
derive from bosonic or pair degrees of freedom; they
have an onset temperature  which we define as
$T_\text{fluc} = \Tc + \delta \Tc$.
This is expected to be significantly below the pseudogap
onset $T^*$. At this latter temperature, a gap in the fermionic excitation spectrum
first starts to appear, reflecting the onset of pair formation.
That $T_\text{fluc}$
and $T^*$ are distinct temperatures is a consequence of the fact that
there must be an appreciable number of pairs before they are clearly observable in
thermodynamical properties and transport. 

For the case of conventional fluctuations,
$T_\text{fluc}$
can be associated with the characteristic size of the critical region,
which can be related to
$G_i$, the Ginzburg-Levanyuk number. This is, of
course, extremely small in 3D although somewhat larger in 2D. (For conventional
  fluctuation theory~\cite{VarlamovBook}, in 3D, $\delta \Tc/T_{c0}
  \approx \sqrt{G_i}$, with $G_i \sim 80(\Tc/\EF)^4$. In 2D, $ \delta
  \Tc/T_{c0} \approx G_i \ln 1/G_i$, with $G_i \approx
  \Tc/\EF$. Here,
  $T_{c0}$ is the mean-field transition temperature.) 

In the
crossover scenario one can address this somewhat different fluctuation
picture in a more quantitative fashion.  The onset temperature
($T_{\text{fluc}}$) for pair-fluctuation effects on thermodynamics and
transport requires sufficiently small but
non-vanishing~\cite{Boyack2021} values for the pair chemical
potential, $|\mu_\text{pair}|$.  In this way, $ \delta
\Tc$ represents the temperature range over which non-condensed
  pairs are present in moderate quantity. Our discussion in this
section has thus emphasized that the onset temperature for
fluctuations is necessarily distinct, not only from
$\Tc$, but also from $T^*$.

\section{Discussion}

This paper is motivated by the observation that, because there are so
many disparate approaches to understanding high-temperature cuprate
superconductivity with as yet no consensus, for future progress it is
important to subject candidate theories to falsifiability tests as
much as is possible.  Here we address one particular scenario: the
BCS-BEC crossover picture.  This has an added advantage among cuprate
theories of being experimentally realized both in Fermi gas
superfluids~\cite{Chen2005,Giorgini2008,Randeria2014} and in a broader
class of strongly correlated superconductors~\cite{Chen2022},
which include some organic superconductors, twisted graphene families,
interfacial superconductors and gated superconducting devices.  These
systems provide an instructive knowledge base for what to expect with
different experimental probes.  Importantly, with this knowledge base
we have learned how to address the applicability of
crossover theory~\cite{Chen2022}.

In this paper, we have argued that the GL coherence length
$\xi_0^{\coh}$ is a preferred parameter for assessing the
appropriateness of BCS-BEC crossover theory for the cuprate family,
when it is measured systematically across the $\Tc$ dome in the phase
diagram and for the different cuprate families. We emphasize that this
coherence length corresponds to temperatures around and slightly above
$\Tc$ as it is associated with normal-state pairs.  This is
necessarily different from the size of Cooper pairs and also from the BCS expression for the zero-temperature coherence length:
$\xi_0^\text{BCS} = \hbar \vF /\left(\pi \Delta_0\right)$ where $\vF$ is the Fermi
velocity and $\Delta_0=\Delta^\text{BCS}(T=0)$. That the behavior is different from BCS theory should be
obvious as BCS theory does not contain preformed pairs.  Note also
that the coherence length extracted at $T \gtrsim \Tc$ in the cuprates
should also not be confused with a counterpart measured in the ground
state which has very different properties, relating to the superconducting condensate.

This same GL coherence length has been extensively studied in other
BCS-BEC crossover candidate systems. Indeed, one can see by comparing
the behavior for the organic superconductor in Fig.~\ref{fig:1}(a)
with the prediction in Fig.~\ref{fig:1}(b) for the cuprates that these
plots are rather similar, although the horizontal axes represent
different variables.  For the cuprates, one sees that the GL coherence
length is predicted to monotonically decrease with increased
underdoping, which reflects the fact that the pairing strength is
strongest in the most underdoped systems.  Also predicted in
Fig.~\ref{fig:1}(b) is that the minimum value of the coherence length
will not be as short as for the organic family, which seems to suggest
these latter systems are closer to the BEC regime.

It is important to note that for assessing the appropriateness of a
BEC scenario where fermions are absent, there are more direct
experiments. Rather than focusing on the coherence length, one can
study the chemical potential to determine whether, as would be
expected, all signs of a Fermi surface have disappeared. Using this
approach, recent work~\cite{Sous2023} has demonstrated that the
cuprates are nowhere near the BEC endpoint of the crossover, where the
chemical potential approaches the band bottom. 
Notably, however, the failure to observe BEC signatures does not constitute evidence for
the `the absence of BCS-BEC crossover'.

That in the cuprates we are lacking a systematic characterization of
the GL coherence length $\xicoh$, over the entire class of cuprate
superconductors, is perhaps surprising, as it is one of the most
fundamental properties of any superconductor.  Moreover, with very few
exceptions, detailed measurements of $\xicoh$ have been used to
provide support for or against a BCS-BEC crossover scenario in nearly
all other candidate superconductors that have been
studied~\cite{Chen2022}. This serves to emphasize how central a role $\xicoh$ has
played, and the importance of further experiments on cuprates. In the process, these
types of experiments will clarify the relevance (or lack thereof) of the
BCS-BEC crossover scenario for the high-transition temperature
copper-oxide superconductors.

\section{Methods}

\subsection{Theory underlying BCS-BEC crossover}

To determine the GL coherence length within BCS-BEC crossover theory, 
it is useful to summarize a few simple equations.
We adopt the particular version of
BCS-BEC crossover theory which builds on the $T=0$ BCS ground state,
\begin{equation}
  \Psi^{\textrm{BCS}}=\Pi_{\vect{k}}
  \left(\uk+\vk a_{\vect{k},\uparrow}^{\dagger} a_{-\vect{k},\downarrow}^{\dagger}\right)|0\rangle.
\label{eq:1}
\end{equation}
This state, originally devised for weak-coupling, can be readily generalized~\cite{Leggett1980}
to incorporate stronger pairing glue
through a self-consistent calculation of the parameters $\uk$ and $\vk$, which can be determined in conjunction
with the fermionic chemical potential $\mu$ as the pairing interaction is varied.

The coherence length, which appears in Eq.~\eqref{eq:4}, depends on the pair density $n_\pair$ and pair mass $M_\pair$. 
These two quantities are important for arriving at the plots in Fig.~\ref{fig:1}(b)
and Fig.~\ref{fig:PhaseDiagram}. 
They must be determined self consistently and we do so
here using a particular theory~\cite{Chen2005,Chen2022},
designed to be consistent with Eq.~\eqref{eq:1} and its finite-temperature extension,
as established by Kadanoff and Martin~\cite{Kadanoff1961}. 
Within this theory one can show that Eq.~\eqref{eq:2} is equivalent to a generalized Thouless condition, which dictates that
the bosonic chemical potential of preformed pairs, $\mu_\pair$, which enters into their 
propagator (called the $t$-matrix) must vanish at $T = \Tc$. 
This generalized Thouless condition will, in turn, lead to a BCS-like gap equation (for $T$ at $\Tc$),
\begin{equation}
1= (- U) \sum_{\vect k} \frac{1 - 2f(E_{\vect k})}{2 E_{\vect k}}  \varphi_{\vect k}^2 \bigg\vert_{T=\Tc},
\label{eq:3}
\end{equation}
where $f(x)=\left[\exp\left(x/(\kB T)\right) +1\right]^{-1}$ is the Fermi-Dirac distribution function, $E_{\vect k} = \sqrt{\xi_{\vect k}^2 + |\Delta(\Tc)\varphi_{\vect{k}}|^2}$ with
$\xi_{\vect k}=\epsilon_{\vect k} -\mu$, and $\varphi_{\vect{k}} = \cos k_x - \cos k_y$ is the $d$-wave pairing symmetry form
factor. Here, $-U>0$ represents the strength of the attractive interaction. Note that the central change from strict BCS theory
(aside from a self-consistent readjustment of the fermionic chemical potential) is that $\Tc$ is determined in the presence of a nonzero excitation gap, $\Delta(\Tc)$, reflecting the non-condensed pairs.

The process of establishing 
Eq.~\eqref{eq:3}
provides values for 
$n_\pair$ and $M_\pair$ associated with our extended form of BCS theory
having a ground state of the form
Eq.~\eqref{eq:1}.
While Thouless has argued that a divergence of a sum of `ladder'
diagrams (within a pair propagator) is to be associated with the
BCS transition temperature,
Kadanoff and Martin established that this Thouless
condition can be extended to characterize the full
BCS temperature-dependent gap equation for all $T\leq T_\text{c}$, provided one
adopts a particular form for the  
pair propagator or $t$-matrix 
\begin{equation}
\label{eq:12}
\frac{1}{t(i\Omega_m, \vecq)}= T\sum_n \sum_{\veck} G(i\omega_n,\veck)G_{0}(i\Omega_m - i\omega_n, \vecq -\veck)+\frac{1}{U},
\end{equation}
The bare and dressed fermionic Green's functions 
in the above equation
are respectively $G_0(i\omega_n,\veck)=\left(i\omega_n - \xi_{\veck}\right)^{-1}$ and
$G(i\omega_n, \veck) \equiv \left[ G_0^{-1}(i\omega_n, \veck) - \Sigma(i\omega_n,\veck) \right]^{-1}$,
with $\Sigma(i\omega_n,\veck) =- \Delta^2 G_0(-i\omega_n, -\veck)$. 
$\hbar \omega_n=(2n+1)\pi \kB T$ and $\hbar \Omega_m =2 m \pi  \kB T$ are fermionic and bosonic Matsubara frequencies (times $\hbar$), respectively. 

\subsection{Calculation of pair mass and number density}

We are now in a position to compute the pair mass and number
density from $t(i\Omega_m,\vecq)$. After analytical continuation,
$i\Omega_m \rightarrow \Omega + i 0^+$, we expand the (inverse)
$t$-matrix for small argument $\Omega$ and $\vecq$ to find
\begin{equation}
t(\Omega, \vecq) \approx  \frac{Z^{-1}}{\Omega - \Omega_{\vecq} +\mupair},
\label{eq:15}
\end{equation}
where the pair mass can be calculated from the pair dispersion
$\Omega_{\mathbf{q}} = \hbar^2\mathbf{q}^2/ (2 M_{\pair})$. 
In this equation $Z$ is a constant independent of $\Omega$ and $\vecq$. 
$\{ M_{\pair}, \mupair, Z\}$ are all functions of the fermionic gap $\Delta$ and chemical potential $\mu$,
which are in turn functions of $|U|$ and temperature $T$ for given total carrier density $n$. 
Finally, one can obtain the density of non-condensed pairs by treating them as stable and independent bosons,
for which we have
\begin{equation}
n_\pair =\sum_{\mathbf{q}} b(\Omega_{\mathbf{q}} -\mupair) = Z \Delta^2. 
\label{eq:16}
\end{equation}
Here, $b(x)=[\exp\left(x/(\kB T)\right)-1]^{-1}$ is the Bose-Einstein distribution function.
To derive the last equality in Eq.~\eqref{eq:16}, we have used $\Delta^2 = -T\sum_m \sum_\vecq t(i\Omega_m,\vecq)$. 
Equation~\eqref{eq:16} is valid for $T\ge \Tc$, while for $T < \Tc$, where $\mupair \equiv 0$, the $\vecq=0$ component, which represents condensed pairs, needs
to be treated separately.  

Right at $T=\Tc$ and for given $\{|U|, n\}$, we solve the gap equation Eq.~\eqref{eq:3} and Eq.~\eqref{eq:16} with $\mupair=0$, together with the total electron density constraint 
\begin{equation}
n= \sum_\veck  \left[1- \frac{\xi_\veck}{E_\veck} \tanh\left(\frac{E_\veck}{2 \kB T}\right)\right],
\end{equation}
to determine $\{\Tc, \Delta(\Tc),\mu(\Tc)\}$. 
In this way we can map out the $\Tc -|U|$ phase diagram for a given density $n$. The result is schematically shown in Fig.~\ref{fig:PhaseDiagram},
where the pseudogap onset temperature $T^*$ is obtained by solving the mean-field BCS $T_c$ equation in the absence of non-condensed pairs. 
Furthermore, from the calculated $\mu$ and $\Delta$ we can compute $\{ M_\pair, \mu_\pair\}$ using Eq.~\eqref{eq:15}.
Then substituting the results into Eq.~\eqref{eq:4} gives us $\xi_0^\coh$ as a function of $|U|$ and $n$. 
In application to cuprate superconductors, we use the calculated $T^*/\Tc$ ratio to determine $|U|$ for given hole doping $p=1-n$, by following the fitting procedure outlined in the Section `Application to the cuprates'.
This allows us to determine $\xi_0^\coh$ as a function of hole doping $p$ for the entire $\Tc$ dome as shown in Fig.~\ref{fig:1}(b). 

\section{Data Availability}
The data analyzed in the current study are available from the author Qijin Chen on reasonable request. 

\section{Code Availability}
The codes used for the current study are available from the author Qijin Chen on reasonable request. 

\section{Acknowledgments} 
We thank Steve Kivelson, John Sous, and Yu He for stimulating discussions.
We also thank Yayu Wang for discussions on related experiments. 
This work was partially (K. L., Z. W.) supported by the Department of Energy (DE-SC0019216). Q. C. was supported by the Innovation Program for Quantum Science and Technology (Grant No. 2021ZD0301904). R. B. was supported by the Department of Physics and Astronomy, Dartmouth College. 

\section{Competing Interests}
The authors declare no competing interests. 

\section{Author Contributions}
K. L. conceived and supervised the project. Q. C. performed the computations. Q. C. and Z. W. contributed to the acquisition of the data and preparation of figures. 
All authors have contributed to the interpretation of the data and the drafting as well as the revision of the manuscript. 

\section{Additional Information}
Correspondence and requests for materials should be addressed to the authors Qijin Chen and K. Levin. 


\begin{thebibliography}{10}
\expandafter\ifx\csname url\endcsname\relax
  \def\url#1{\texttt{#1}}\fi
\expandafter\ifx\csname urlprefix\endcsname\relax\def\urlprefix{URL }\fi
\providecommand{\bibinfo}[2]{#2}
\providecommand{\eprint}[2][]{\url{#2}}

\bibitem{Lee2006}
\bibinfo{author}{Lee, P.~A.}, \bibinfo{author}{Nagaosa, N.} \&
  \bibinfo{author}{Wen, X.-G.}
\newblock \bibinfo{title}{Doping a mott insulator: Physics of high-temperature
  superconductivity}.
\newblock \emph{\bibinfo{journal}{\rmp}} \textbf{\bibinfo{volume}{78}},
  \bibinfo{pages}{17} (\bibinfo{year}{2006}).

\bibitem{Chen2005}
\bibinfo{author}{Chen, Q.~J.}, \bibinfo{author}{Stajic, J.},
  \bibinfo{author}{Tan, S.~N.} \& \bibinfo{author}{Levin, K.}
\newblock \bibinfo{title}{{BCS--BEC crossover: From} high temperature
  superconductors to ultracold superfluids}.
\newblock \emph{\bibinfo{journal}{Phys. Rep.}} \textbf{\bibinfo{volume}{412}},
  \bibinfo{pages}{1--88} (\bibinfo{year}{2005}).

\bibitem{Keimer2015}
\bibinfo{author}{Keimer, B.}, \bibinfo{author}{Kivelson, S.~A.},
  \bibinfo{author}{Norman, M.~R.}, \bibinfo{author}{Uchida, S.} \&
  \bibinfo{author}{Zaanen, J.}
\newblock \bibinfo{title}{From quantum matter to high-temperature
  superconductivity in copper oxides}.
\newblock \emph{\bibinfo{journal}{Nature}} \textbf{\bibinfo{volume}{518}},
  \bibinfo{pages}{179--186} (\bibinfo{year}{2015}).

\bibitem{Fradkin2015}
\bibinfo{author}{Fradkin, E.}, \bibinfo{author}{Kivelson, S.~A.} \&
  \bibinfo{author}{Tranquada, J.~M.}
\newblock \bibinfo{title}{Colloquium: Theory of intertwined orders in high
  temperature superconductors}.
\newblock \emph{\bibinfo{journal}{\rmp}} \textbf{\bibinfo{volume}{87}},
  \bibinfo{pages}{457} (\bibinfo{year}{2015}).

\bibitem{Harrison2022}
\bibinfo{author}{Harrison, N.} \& \bibinfo{author}{Chan, M.~K.}
\newblock \bibinfo{title}{Magic gap ratio for optimally robust fermionic
  condensation and its implications for high $\text{\ensuremath{-}}{T}_{c}$
  superconductivity}.
\newblock \emph{\bibinfo{journal}{Phys. Rev. Lett.}}
  \textbf{\bibinfo{volume}{129}}, \bibinfo{pages}{017001}
  (\bibinfo{year}{2022}).

\bibitem{Sous2023}
\bibinfo{author}{Sous, J.}, \bibinfo{author}{He, Y.} \&
  \bibinfo{author}{Kivelson, S.~A.}
\newblock \bibinfo{title}{Absence of a {BCS-BEC} crossover in the cuprate
  superconductors}.
\newblock \emph{\bibinfo{journal}{npj Quantum Mater.}}
  \textbf{\bibinfo{volume}{8}}, \bibinfo{pages}{25} (\bibinfo{year}{2023}).

\bibitem{Kasahara2016}
\bibinfo{author}{Kasahara, S.} \emph{et~al.}
\newblock \bibinfo{title}{Giant superconducting fluctuations in the compensated
  semimetal {FeSe at the BCS--BEC} crossover}.
\newblock \emph{\bibinfo{journal}{Nat. Commun.}} \textbf{\bibinfo{volume}{7}},
  \bibinfo{pages}{12843} (\bibinfo{year}{2016}).

\bibitem{Kasahara2014}
\bibinfo{author}{Kasahara, S.} \emph{et~al.}
\newblock \bibinfo{title}{Field-induced superconducting phase of {FeSe in the
  BCS-BEC} cross-over}.
\newblock \emph{\bibinfo{journal}{Proc. Nat'l Acad. Sci. U.S.A.}}
  \textbf{\bibinfo{volume}{111}}, \bibinfo{pages}{16309--16313}
  (\bibinfo{year}{2014}).

\bibitem{Okazaki2014}
\bibinfo{author}{Okazaki, K.} \emph{et~al.}
\newblock \bibinfo{title}{Superconductivity in an electron band just above the
  {Fermi level: possible route to BCS-BEC} superconductivity}.
\newblock \emph{\bibinfo{journal}{Sci. Rep.}} \textbf{\bibinfo{volume}{4}},
  \bibinfo{pages}{4109} (\bibinfo{year}{2014}).

\bibitem{Mizukami2021}
\bibinfo{author}{Mizukami, Y.} \emph{et~al.}
\newblock \bibinfo{title}{Thermodynamics of transition to {BCS-BEC crossover
  superconductivity in FeSe$_{1-x}$S$_x$}}.
\newblock \emph{\bibinfo{journal}{Commun. Phys.}} \textbf{\bibinfo{volume}{6}},
  \bibinfo{pages}{183} (\bibinfo{year}{2023}).

\bibitem{Hanaguri2019}
\bibinfo{author}{Hanaguri, T.} \emph{et~al.}
\newblock \bibinfo{title}{Quantum vortex core and missing pseudogap in the
  multiband {BCS-BEC} crossover superconductor {FeSe}}.
\newblock \emph{\bibinfo{journal}{Phys. Rev. Lett.}}
  \textbf{\bibinfo{volume}{122}}, \bibinfo{pages}{077001}
  (\bibinfo{year}{2019}).

\bibitem{Shibauchi2020}
\bibinfo{author}{Shibauchi, T.}, \bibinfo{author}{Hanaguri, T.} \&
  \bibinfo{author}{Matsuda, Y.}
\newblock \bibinfo{title}{Exotic superconducting states in {FeSe-based}
  materials}.
\newblock \emph{\bibinfo{journal}{J. Phys. Soc. Jpn.}}
  \textbf{\bibinfo{volume}{89}}, \bibinfo{pages}{102002}
  (\bibinfo{year}{2020}).

\bibitem{Kang2020}
\bibinfo{author}{Kang, B.~L.} \emph{et~al.}
\newblock \bibinfo{title}{{Preformed Cooper Pairs in Layered FeSe-Based
  Superconductors}}.
\newblock \emph{\bibinfo{journal}{Phys. Rev. Lett.}}
  \textbf{\bibinfo{volume}{125}}, \bibinfo{pages}{097003}
  (\bibinfo{year}{2020}).

\bibitem{Faeth2021}
\bibinfo{author}{Faeth, B.~D.} \emph{et~al.}
\newblock \bibinfo{title}{Incoherent {Cooper} pairing and pseudogap behavior in
  single-layer {$\mathrm{FeSe}/\mathrm{SrTi}{\mathrm{O}}_{3}$}}.
\newblock \emph{\bibinfo{journal}{Phys. Rev. X}} \textbf{\bibinfo{volume}{11}},
  \bibinfo{pages}{021054} (\bibinfo{year}{2021}).

\bibitem{Mckenzie1997}
\bibinfo{author}{McKenzie, R.~H.}
\newblock \bibinfo{title}{Similarities between organic and cuprate
  superconductors}.
\newblock \emph{\bibinfo{journal}{Science}} \textbf{\bibinfo{volume}{278}},
  \bibinfo{pages}{820--821} (\bibinfo{year}{1997}).

\bibitem{Imajo2021}
\bibinfo{author}{Imajo, S.} \emph{et~al.}
\newblock \bibinfo{title}{Extraordinary $\pi$-electron superconductivity
  emerging from a quantum spin liquid}.
\newblock \emph{\bibinfo{journal}{Phys. Rev. Research}}
  \textbf{\bibinfo{volume}{3}}, \bibinfo{pages}{033026} (\bibinfo{year}{2021}).

\bibitem{Matsumura2022}
\bibinfo{author}{Matsumura, Y.}, \bibinfo{author}{Yamashita, S.},
  \bibinfo{author}{Akutsu, H.} \& \bibinfo{author}{Nakazawa, Y.}
\newblock \bibinfo{title}{Thermodynamic measurements of doped dimer-{Mott}
  organic superconductor under pressure}.
\newblock \emph{\bibinfo{journal}{Low Temp. Phys.}}
  \textbf{\bibinfo{volume}{48}}, \bibinfo{pages}{51--56}
  (\bibinfo{year}{2022}).

\bibitem{Oike2017}
\bibinfo{author}{Oike, H.} \emph{et~al.}
\newblock \bibinfo{title}{Anomalous metallic behaviour in the doped spin liquid
  candidate {$\kappa$-(ET)$_4$Hg$_{2.89}$Br$_8$}}.
\newblock \emph{\bibinfo{journal}{Nat. commun.}} \textbf{\bibinfo{volume}{8}},
  \bibinfo{pages}{756} (\bibinfo{year}{2017}).

\bibitem{Suzuki2022}
\bibinfo{author}{Suzuki, Y.} \emph{et~al.}
\newblock \bibinfo{title}{Mott-driven {BEC-BCS} crossover in a doped spin
  liquid candidate {$\kappa$-(BEDT- TTF)$_4$Hg$_{2.89}$Br$_8$}}.
\newblock \emph{\bibinfo{journal}{Phys. Rev. X}} \textbf{\bibinfo{volume}{12}},
  \bibinfo{pages}{011016} (\bibinfo{year}{2022}).

\bibitem{Cao2018}
\bibinfo{author}{Cao, Y.} \emph{et~al.}
\newblock \bibinfo{title}{Unconventional superconductivity in magic-angle
  graphene superlattices}.
\newblock \emph{\bibinfo{journal}{Nature}} \textbf{\bibinfo{volume}{556}},
  \bibinfo{pages}{43--50} (\bibinfo{year}{2018}).

\bibitem{Oh2021}
\bibinfo{author}{Oh, M.} \emph{et~al.}
\newblock \bibinfo{title}{Evidence for unconventional superconductivity in
  twisted bilayer graphene}.
\newblock \emph{\bibinfo{journal}{Nature}} \textbf{\bibinfo{volume}{600}},
  \bibinfo{pages}{240--245} (\bibinfo{year}{2021}).

\bibitem{Park2021}
\bibinfo{author}{Park, J.~M.}, \bibinfo{author}{Cao, Y.},
  \bibinfo{author}{Watanabe, K.}, \bibinfo{author}{Taniguchi, T.} \&
  \bibinfo{author}{Jarillo-Herrero, P.}
\newblock \bibinfo{title}{Tunable strongly coupled superconductivity in
  magic-angle twisted trilayer graphene}.
\newblock \emph{\bibinfo{journal}{Nature}} \textbf{\bibinfo{volume}{590}},
  \bibinfo{pages}{249--255} (\bibinfo{year}{2021}).

\bibitem{Kim2022}
\bibinfo{author}{Kim, H.} \emph{et~al.}
\newblock \bibinfo{title}{Evidence for unconventional superconductivity in
  twisted trilayer graphene}.
\newblock \emph{\bibinfo{journal}{Nature}} \textbf{\bibinfo{volume}{606}},
  \bibinfo{pages}{494--500} (\bibinfo{year}{2022}).

\bibitem{Nakagawa2021}
\bibinfo{author}{Nakagawa, Y.} \emph{et~al.}
\newblock \bibinfo{title}{Gate-controlled {BCS-BEC} crossover in a
  two-dimensional superconductor}.
\newblock \emph{\bibinfo{journal}{Science}} \textbf{\bibinfo{volume}{372}},
  \bibinfo{pages}{190--195} (\bibinfo{year}{2021}).

\bibitem{Saito2016}
\bibinfo{author}{Saito, Y.}, \bibinfo{author}{Nojima, T.} \&
  \bibinfo{author}{Iwasa, Y.}
\newblock \bibinfo{title}{Highly crystalline {2D} superconductors}.
\newblock \emph{\bibinfo{journal}{Nat. Rev. Mater.}}
  \textbf{\bibinfo{volume}{2}}, \bibinfo{pages}{16094} (\bibinfo{year}{2016}).

\bibitem{Nakagawa2018}
\bibinfo{author}{Nakagawa, Y.} \emph{et~al.}
\newblock \bibinfo{title}{Gate-controlled low carrier density superconductors:
  {Toward the two-dimensional BCS-BEC} crossover}.
\newblock \emph{\bibinfo{journal}{Phys. Rev. B}} \textbf{\bibinfo{volume}{98}},
  \bibinfo{pages}{064512} (\bibinfo{year}{2018}).

\bibitem{Richter2013}
\bibinfo{author}{Richter, C.} \emph{et~al.}
\newblock \bibinfo{title}{Interface superconductor with gap behaviour like a
  high-temperature superconductor}.
\newblock \emph{\bibinfo{journal}{Nature}} \textbf{\bibinfo{volume}{502}},
  \bibinfo{pages}{528--531} (\bibinfo{year}{2013}).

\bibitem{Bozovic2020}
\bibinfo{author}{Bo{\v{z}}ovi{\'c}, I.} \& \bibinfo{author}{Levy, J.}
\newblock \bibinfo{title}{Pre-formed Cooper pairs in copper oxides and
  {LaAlO$_3$/SrTiO$_3$} heterostructures}.
\newblock \emph{\bibinfo{journal}{Nat. Phys.}} \textbf{\bibinfo{volume}{16}},
  \bibinfo{pages}{712--717} (\bibinfo{year}{2020}).

\bibitem{Cheng2015}
\bibinfo{author}{Cheng, G.} \emph{et~al.}
\newblock \bibinfo{title}{Electron pairing without superconductivity}.
\newblock \emph{\bibinfo{journal}{Nature}} \textbf{\bibinfo{volume}{521}},
  \bibinfo{pages}{196--199} (\bibinfo{year}{2015}).

\bibitem{Liu2022}
\bibinfo{author}{Liu, X.} \emph{et~al.}
\newblock \bibinfo{title}{Crossover between strongly coupled and weakly coupled
  exciton superfluids}.
\newblock \emph{\bibinfo{journal}{Science}} \textbf{\bibinfo{volume}{375}},
  \bibinfo{pages}{205--209} (\bibinfo{year}{2022}).

\bibitem{Leggett1980}
\bibinfo{author}{Leggett, A.~J.}
\newblock \bibinfo{title}{Diatomic molecules and {Cooper} pairs}.
\newblock In \bibinfo{editor}{Pekalski, A.} \& \bibinfo{editor}{Przystawa,
  J.~A.} (eds.) \emph{\bibinfo{booktitle}{Modern Trends in the Theory of
  Condensed Matter}}, vol. \bibinfo{volume}{115} of
  \emph{\bibinfo{series}{Lecture Notes in Physics}}, \bibinfo{pages}{13--27}
  (\bibinfo{publisher}{Springer-Verlag}, \bibinfo{address}{Berlin, West
  Germany}, \bibinfo{year}{1980}).
\newblock \bibinfo{note}{Proceedings of the XVI Karpacz Winter School of
  Theoretical Physics, February 19 - March 3, 1979, Karpacz, Poland}.

\bibitem{Eagles1969}
\bibinfo{author}{Eagles, D.~M.}
\newblock \bibinfo{title}{Possible pairing without superconductivity at low
  carrier concentrations in bulk and thin-film superconducting semiconductors}.
\newblock \emph{\bibinfo{journal}{Phys. Rev.}} \textbf{\bibinfo{volume}{186}},
  \bibinfo{pages}{456--463} (\bibinfo{year}{1969}).

\bibitem{Nozieres1985}
\bibinfo{author}{Nozi\`{e}res, P.} \& \bibinfo{author}{Schmitt-Rink, S.}
\newblock \bibinfo{title}{Bose condensation in an attractive fermion gas: from
  weak to strong coupling superconductivity}.
\newblock \emph{\bibinfo{journal}{J. Low Temp. Phys.}}
  \textbf{\bibinfo{volume}{59}}, \bibinfo{pages}{195--211}
  (\bibinfo{year}{1985}).

\bibitem{Giorgini2008}
\bibinfo{author}{Giorgini, S.}, \bibinfo{author}{Pitaevskii, L.~P.} \&
  \bibinfo{author}{Stringari, S.}
\newblock \bibinfo{title}{Theory of ultracold atomic {Fermi} gases}.
\newblock \emph{\bibinfo{journal}{Rev. Mod. Phys.}}
  \textbf{\bibinfo{volume}{80}}, \bibinfo{pages}{1215--1274}
  (\bibinfo{year}{2008}).

\bibitem{Randeria2014}
\bibinfo{author}{Randeria, M.} \& \bibinfo{author}{Taylor, E.}
\newblock \bibinfo{title}{Crossover from {Bardeen-Cooper-Schrieffer to
  Bose-Einstein condensation and the unitary Fermi gas}}.
\newblock \emph{\bibinfo{journal}{Annu. Rev. Condens. Matter Phys.}}
  \textbf{\bibinfo{volume}{5}}, \bibinfo{pages}{209--232}
  (\bibinfo{year}{2014}).

\bibitem{Chen2022}
\bibinfo{author}{Chen, Q.~J.}, \bibinfo{author}{Wang, Z.~Q.},
  \bibinfo{author}{Boyack, R.}, \bibinfo{author}{Yang, S.~L.} \&
  \bibinfo{author}{Levin, K.}
\newblock \bibinfo{title}{When superconductivity crosses over: From {BCS} to
  {BEC}}.
\newblock \emph{\bibinfo{journal}{arXiv:2208.01774}}  (\bibinfo{year}{2022}).

\bibitem{Leggett2006}
\bibinfo{author}{Leggett, A.~J.}
\newblock \bibinfo{title}{What do we know about high {Tc}?}
\newblock \emph{\bibinfo{journal}{Nat. Phys.}} \textbf{\bibinfo{volume}{2}},
  \bibinfo{pages}{134--136} (\bibinfo{year}{2006}).

\bibitem{Chand2012}
\bibinfo{author}{Chand, M.} \emph{et~al.}
\newblock \bibinfo{title}{Phase diagram of the strongly disordered $s$-wave
  superconductor {NbN} close to the metal-insulator transition}.
\newblock \emph{\bibinfo{journal}{Phys. Rev. B}} \textbf{\bibinfo{volume}{85}},
  \bibinfo{pages}{014508} (\bibinfo{year}{2012}).

\bibitem{Suzuki1991}
\bibinfo{author}{Suzuki, M.} \& \bibinfo{author}{Hikita, M.}
\newblock \bibinfo{title}{Resistive transition, magnetoresistance, and
  anisotropy in
  {${\mathrm{La}}_{2\mathrm{\ensuremath{-}}\mathit{x}}$${\mathrm{Sr}}_{\mathit{x}}$${\mathrm{CuO}}_{4}$}
  single-crystal thin films}.
\newblock \emph{\bibinfo{journal}{Phys. Rev. B}} \textbf{\bibinfo{volume}{44}},
  \bibinfo{pages}{249--261} (\bibinfo{year}{1991}).

\bibitem{VarlamovBook}
\bibinfo{author}{Larkin, A.~I.} \& \bibinfo{author}{Varlamov, A.~A.}
\newblock \emph{\bibinfo{title}{Theory of Fluctuations in Superconductors}}.
\newblock International Series of Monographs on Physics
  (\bibinfo{publisher}{OUP Oxford}, \bibinfo{year}{2009}).

\bibitem{Varlamov2018}
\bibinfo{author}{Varlamov, A.~A.}, \bibinfo{author}{Galda, A.} \&
  \bibinfo{author}{Glatz, A.}
\newblock \bibinfo{title}{Fluctuation spectroscopy: From rayleigh-jeans waves
  to abrikosov vortex clusters}.
\newblock \emph{\bibinfo{journal}{Rev. Mod. Phys.}}
  \textbf{\bibinfo{volume}{90}}, \bibinfo{pages}{015009}
  (\bibinfo{year}{2018}).

\bibitem{Patton1971}
\bibinfo{author}{Patton, B.~R.}
\newblock \bibinfo{title}{Fluctuation theory of the superconducting transition
  in restricted dimensionality}.
\newblock \emph{\bibinfo{journal}{Phys. Rev. Lett.}}
  \textbf{\bibinfo{volume}{27}}, \bibinfo{pages}{1273--1276}
  (\bibinfo{year}{1971}).

\bibitem{Ullah1991}
\bibinfo{author}{Ullah, S.} \& \bibinfo{author}{Dorsey, A.~T.}
\newblock \bibinfo{title}{Effect of fluctuations on the transport properties of
  type-ii superconductors in a magnetic field}.
\newblock \emph{\bibinfo{journal}{\prb}} \textbf{\bibinfo{volume}{44}},
  \bibinfo{pages}{262} (\bibinfo{year}{1991}).

\bibitem{Boyack2018}
\bibinfo{author}{Boyack, R.}, \bibinfo{author}{Chen, Q.~J.},
  \bibinfo{author}{Varlamov, A.~A.} \& \bibinfo{author}{Levin, K.}
\newblock \bibinfo{title}{Cuprate diamagnetism in the presence of a pseudogap:
  {Beyond} the standard fluctuation formalism}.
\newblock \emph{\bibinfo{journal}{Phys. Rev. B}} \textbf{\bibinfo{volume}{97}},
  \bibinfo{pages}{064503} (\bibinfo{year}{2018}).

\bibitem{Boyack2019}
\bibinfo{author}{Boyack, R.}, \bibinfo{author}{Wang, X.~Y.},
  \bibinfo{author}{Chen, Q.~J.} \& \bibinfo{author}{Levin, K.}
\newblock \bibinfo{title}{Combined effects of pairing fluctuations and a
  pseudogap in the cuprate {Hall} coefficient}.
\newblock \emph{\bibinfo{journal}{Phys. Rev. B}} \textbf{\bibinfo{volume}{99}},
  \bibinfo{pages}{134504} (\bibinfo{year}{2019}).

\bibitem{Kadanoff1961}
\bibinfo{author}{Kadanoff, L.~P.} \& \bibinfo{author}{Martin, P.~C.}
\newblock \bibinfo{title}{Theory of many-particle systems. {II.
  Superconductivity}}.
\newblock \emph{\bibinfo{journal}{Phys. Rev.}} \textbf{\bibinfo{volume}{124}},
  \bibinfo{pages}{670--697} (\bibinfo{year}{1961}).

\bibitem{Chen1999}
\bibinfo{author}{Chen, Q.~J.}, \bibinfo{author}{Kosztin, I.},
  \bibinfo{author}{Jank\'o, B.} \& \bibinfo{author}{Levin, K.}
\newblock \bibinfo{title}{Superconducting transitions from the pseudogap state:
  $d$-wave symmetry, lattice, and low-dimensional effects}.
\newblock \emph{\bibinfo{journal}{Phys. Rev. B}} \textbf{\bibinfo{volume}{59}},
  \bibinfo{pages}{7083--7093} (\bibinfo{year}{1999}).

\bibitem{Wen2003}
\bibinfo{author}{Wen, H.~H.} \emph{et~al.}
\newblock \bibinfo{title}{Hole doping dependence of the coherence length in
  {La$_{2-x}$Sr$_x$CuO$_4$} thin films}.
\newblock \emph{\bibinfo{journal}{Europhys. Lett.}}
  \textbf{\bibinfo{volume}{64}}, \bibinfo{pages}{790} (\bibinfo{year}{2003}).

\bibitem{Sonier2007}
\bibinfo{author}{Sonier, J.~E.} \emph{et~al.}
\newblock \bibinfo{title}{Hole-doping dependence of the magnetic penetration
  depth and vortex core size in
  {$\mathrm{Y}{\mathrm{Ba}}_{2}{\mathrm{Cu}}_{3}{\mathrm{O}}_{y}$}: Evidence
  for stripe correlations near {$\frac{1}{8}$} hole doping}.
\newblock \emph{\bibinfo{journal}{Phys. Rev. B}} \textbf{\bibinfo{volume}{76}},
  \bibinfo{pages}{134518} (\bibinfo{year}{2007}).

\bibitem{Ando2002}
\bibinfo{author}{Ando, Y.} \& \bibinfo{author}{Segawa, K.}
\newblock \bibinfo{title}{Magnetotransport properties of untwinned
  {YBa$_2$Cu$_3$O$_y$ single crystals: novel 60-K-phase} anomalies in the
  charge transport}.
\newblock \emph{\bibinfo{journal}{J. Phys. Chem. Solids}}
  \textbf{\bibinfo{volume}{63}}, \bibinfo{pages}{2253--2257}
  (\bibinfo{year}{2002}).
\newblock \bibinfo{note}{Proceedings of the Conference on Spectroscopies in
  Novel Superconductors}.

\bibitem{Chan2020}
\bibinfo{author}{Chan, M.~K.} \emph{et~al.}
\newblock \bibinfo{title}{Extent of Fermi-surface reconstruction in the
  high-temperature superconductor { HgBa$_2$CuO$_{4+\delta}$}}.
\newblock \emph{\bibinfo{journal}{Proc. Nat'l Acad. Sci. U.S.A.}}
  \textbf{\bibinfo{volume}{117}}, \bibinfo{pages}{9782--9786}
  (\bibinfo{year}{2020}).

\bibitem{Badoux2016}
\bibinfo{author}{Badoux, S.} \emph{et~al.}
\newblock \bibinfo{title}{Critical doping for the onset of {Fermi}-surface
  reconstruction by charge-density-wave order in the cuprate superconductor
  {${\mathrm{La}}_{2\ensuremath{-}x}{\mathrm{Sr}}_{x}{\mathrm{CuO}}_{4}$}}.
\newblock \emph{\bibinfo{journal}{Phys. Rev. X}} \textbf{\bibinfo{volume}{6}},
  \bibinfo{pages}{021004} (\bibinfo{year}{2016}).

\bibitem{Wang2006}
\bibinfo{author}{Wang, Y.}, \bibinfo{author}{Li, L.} \& \bibinfo{author}{Ong,
  N.~P.}
\newblock \bibinfo{title}{Nernst effect in high-${T}_{c}$ superconductors}.
\newblock \emph{\bibinfo{journal}{Phys. Rev. B}} \textbf{\bibinfo{volume}{73}},
  \bibinfo{pages}{024510} (\bibinfo{year}{2006}).

\bibitem{Wang2003}
\bibinfo{author}{Wang, Y.} \emph{et~al.}
\newblock \bibinfo{title}{Dependence of upper critical field and pairing
  strength on doping in cuprates}.
\newblock \emph{\bibinfo{journal}{Science}} \textbf{\bibinfo{volume}{299}},
  \bibinfo{pages}{86--89} (\bibinfo{year}{2003}).

\bibitem{Chang2012}
\bibinfo{author}{Chang, J.} \emph{et~al.}
\newblock \bibinfo{title}{Decrease of upper critical field with underdoping in
  cuprate superconductors}.
\newblock \emph{\bibinfo{journal}{Nat. Phys.}} \textbf{\bibinfo{volume}{8}},
  \bibinfo{pages}{751--756} (\bibinfo{year}{2012}).

\bibitem{Boyack2021}
\bibinfo{author}{Boyack, R.}, \bibinfo{author}{Wang, Z.~Q.},
  \bibinfo{author}{Chen, Q.~J.} \& \bibinfo{author}{Levin, K.}
\newblock \bibinfo{title}{Unified approach to electrical and thermal transport
  in high-{$T_c$} superconductors}.
\newblock \emph{\bibinfo{journal}{\prb}} \textbf{\bibinfo{volume}{104}},
  \bibinfo{pages}{064508} (\bibinfo{year}{2021}).

\bibitem{Vishik2018}
\bibinfo{author}{Vishik, I.~M.}
\newblock \bibinfo{title}{Photoemission perspective on pseudogap,
  superconducting fluctuations, and charge order in cuprates: a review of
  recent progress}.
\newblock \emph{\bibinfo{journal}{Rep. Prog. Phys.}}
  \textbf{\bibinfo{volume}{81}}, \bibinfo{pages}{062501}
  (\bibinfo{year}{2018}).

\end{thebibliography}

\end{document}